\documentclass[11pt]{article}

\usepackage[preprint]{acl}

\usepackage{times}
\usepackage{latexsym}

\usepackage[T1]{fontenc}

\usepackage[utf8]{inputenc}

\usepackage{microtype}

\usepackage{inconsolata}

\usepackage{graphicx}
\usepackage{amsmath}
\usepackage{amssymb}
\usepackage{booktabs}
\usepackage{multirow}
\usepackage{adjustbox}
\usepackage{placeins}
\usepackage{float}
\usepackage{listings}
\usepackage{tikz}

\title{LLM-Enhanced Rumor Detection via Virtual Node Induced Edge Prediction}

\author{
  Jiran Tao\textsuperscript{1} \quad Cheng Wang\textsuperscript{2} \quad
  Binyan Jiang\textsuperscript{1} \\
  {\normalfont\large \textsuperscript{1}Department of Data Science and Artificial Intelligence} \\
  {\normalfont\large Hong Kong Polytechnic University, Hung Hom, Kowloon, Hong Kong} \\
  {\normalfont\large \textsuperscript{2}School of Mathematical Sciences,
  Shanghai Jiao Tong University} \\
  {\normalfont\large Shanghai, China} \\
  {\normalfont\normalsize
  \texttt{22117758r@connect.polyu.hk} \quad
  \texttt{chengwang@sjtu.edu.cn} \quad
  \texttt{by.jiang@polyu.edu.hk}}
}

\begin{document}
\maketitle
\begin{abstract}
The rapid proliferation of rumors on social networks poses a significant threat to information integrity. While rumor dissemination forms complex structural patterns, existing detection methods often fail to capture the intricate interplay between textual coherence and propagation dynamics. Current approaches typically represent nodes through isolated textual embeddings, neglecting the semantic flow across the entire propagation path.  To bridge this gap, we introduce a novel framework that integrates Large Language Models (LLMs) as a structural augmentation layer for graph-based rumor detection. Moving beyond conventional methods, our framework employs LLMs to evaluate information subchains and strategically introduce a virtual node into the graph. This structural modification converts latent semantic patterns into explicit topological features, effectively capturing the textual coherence that has historically been inaccessible to Graph Neural Networks (GNNs). To ensure reliability, we develop a structured prompt framework that mitigates inherent biases in LLMs while maintaining robust graph learning performance. Furthermore, our proposed framework is model-agnostic, meaning it is not constrained to any specific graph learning algorithm or LLMs. Its plug-and-play nature allows for seamless integration with further fine-tuned LLMs and graph techniques in the future, potentially enhancing predictive performance without the need to modify original algorithms.
\end{abstract}

\section{Introduction}
 
While social media’s rapid expansion has revolutionized information sharing, it has simultaneously accelerated the spread of rumors, threatening information credibility and societal stability \citep{ziari2025}. Detecting rumors in social networks is a critical yet challenging task, as rumors typically propagate downward along tree-like graph structures over time, with nodes (representing users or posts) and edges (indicating interactions such as replies or retweets) reflecting the flow of information.
This temporal nature, combined with the semantic complexity of textual data, requires modeling approaches that can effectively capture structural patterns, temporal evolution, and deep semantic content.

Traditional rumor detection methods, while achieving some success, are constrained by notable limitations. Early machine learning approaches that rely on hand-crafted features struggle to adapt to the diversity and noise inherent in social media data \citep{zubiaga2018}. Similarly, conventional GNNs, which are primarily designed to model local graph structures, frequently prove inadequate in capturing the subtle rumor propagation pathways and diffusion patterns evident in subchains of varying lengths. LLMs excel at extracting rich semantic features from text, presenting a promising opportunity to address the semantic shortcomings of traditional GNNs \citep{wang2024}. However, our experiments reveal a key limitation of LLMs when used in isolation for rumor detection: different models exhibit markedly different tendencies due to variations in pre-training data and reinforcement learning objectives. Some LLMs are highly conservative and risk-averse, while others are more aggressive or lenient. Consequently, standalone LLM-based approaches may be inherently unreliable for robust rumor detection, a failure case is shown in Appendix G. 
Given these conflicting behaviors and the inherent complexity of network data, how to effectively utilize LLMs to enhance the performance of rumor detection remains a critical and open question. 
To address these challenges, we propose a novel model-agnostic framework that integrates LLMs and GNNs, leveraging structured subchain propagation patterns to enhance rumor detection accuracy.

Our method utilizes LLMs to analyze information flow within tweet subchains, capturing key propagation patterns while leveraging the extensive internal knowledge. While we employ a structured prompt framework to facilitate the processing of network data and mitigate classification biases of LLMs, our primary innovation lies in the architectural enhancement of the rumor graph. We introduce a virtual node representing the rumor status and augment the existing graph structure using nuanced rumor probabilities generated by LLMs for each subchain. This approach is highly efficient as it relies solely on API calls for the LLMs, eliminating the need for local model tuning or intensive storage infrastructure. The enriched graph is subsequently processed to derive robust node representations, thereby facilitating accurate rumor detection through link prediction between the root node and the virtual node.  
 In summary, our study offers two principal contributions: (1) We propose a pioneering, model-agnostic framework that synergistically integrates LLMs and GNNs for rumor detection, enabling effective fusion of semantic and structural information. (2) We develop an innovative approach to capture unexploited propagation features by restructuring the graph with a virtual node and subchain-based connections, enhancing the detection of complex rumor diffusion patterns.


\section{Related Work}
\subsection{Rumor Detection}
Early rumor detection systems used handcrafted features and classifiers such as SVMs \citep{zubiaga2018}. Subsequent approaches use recurrent networks to model temporal sequences \citep{ma2016}, and model propagation trees using kernels, recursive neural networks, or tree LSTMs \citep{ma2017,ma2018rvnn,kumar2019}.

\subsection{Large Language Models in Fake News Detection}
LLMs have been used as judges for model outputs \citep{zhou2023lima,li2024}. In fact-checking, FactLLaMA combines instruction tuning with external knowledge \citep{cheung2023factllama}, and TrumorGPT retrieves evidence from a health knowledge graph \citep{Hang2025trumorgpt}. LLMs also support rumor detection through influence-guided training sample selection \citep{tian2025llm} and data augmentation \citep{lai2024rumorllm}.

\subsection{Graph Methods in Rumor Detection}
Graph-based rumor detectors model retweet chains and branching reply threads. Bi-GCN captures bidirectional information flow \citep{bian2020}, and propagation GNNs with attention model non-sequential diffusion \citep{wu2020rumor}. Graph adversarial contrastive learning uses adversarial feature transformations \citep{sun2022rumor}, while bipartite user--tweet graphs capture user correlations alongside tree-structured propagation \citep{liu2024rumor}. GSMA extends GraphSAGE with dynamic attention, positional encodings, and sentiment features \citep{hamilton2017inductive,ma2024rumor}.

LLMs also enrich graphs for misinformation detection through topic and entity extraction \citep{ma2024fake}, semantic reshuffling \citep{li2025semantic}, and reasoning over social contexts refined by graph attention \citep{zeng2025}.

In graph learning, master or virtual nodes facilitate message passing and have been studied for link prediction \citep{gilmer2017message,hwang2022virtual}. LLM-GNN uses LLM annotations to supervise a GNN, and TAPE encodes LLM explanations as node features \citep{chen2024labelfree,he2024tape}. GraphEdit uses an LLM to infer node relationships for graph refinement \citep{guo2024graphedit}. LLM-VN uses subchain scores to attach selected replies to a virtual rumor-state node and predicts a link between the source post and that node.






\section{Problem Statement}
\label{gen_inst}
\begin{figure}[t]
  \centering
  \includegraphics[width=0.7\columnwidth]{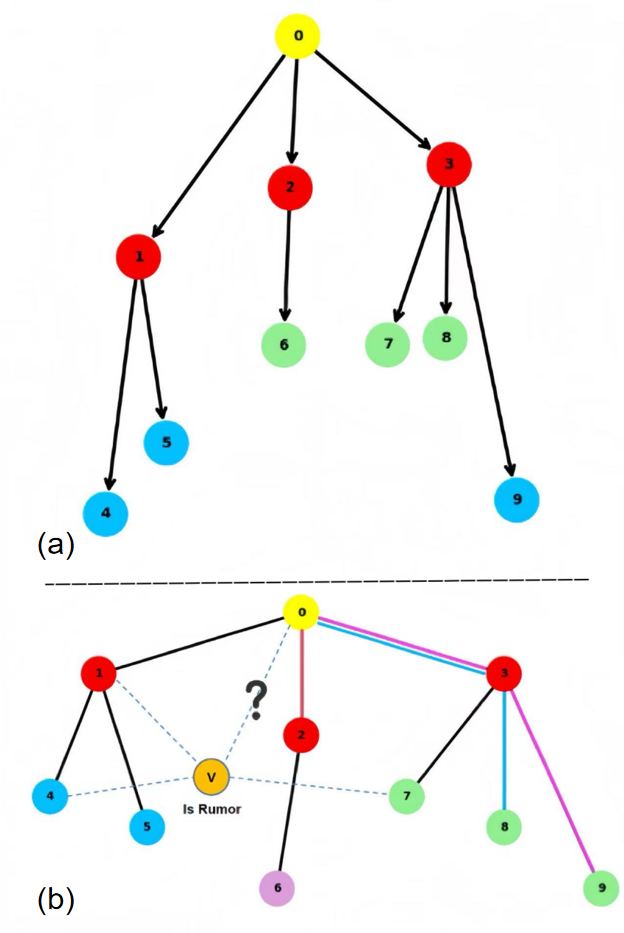}
  \caption{Propagation tree and its augmentation. In (b), colored paths show subchains, and dashed lines connect replies 1, 4, and 7 to the virtual node $v^\ast$. The question mark denotes the link to predict.}
  \label{fig:experiments}
\end{figure}

Let $\mathcal{N} = \{N_1, N_2, \dots, N_s\}$ denote the set of source news articles, where each $N_i \in \mathcal{N}$ represents an individual news item. Associated with each $N_i$ is a set of reactions $\mathcal{R}_i = \{R_{i1}, R_{i2}, \dots, R_{it_i}\}$, forming a multi-level discussion structure beneath each original news post.
Specifically, 1) each news post contains a large number of user replies; 2) these replies may directly respond to the news post or be nested under existing replies as sub-replies, thus forming a tree-like dialogue structure. This tree propagation structure corresponds exactly to the way rumors spread. As shown in Figure~\ref{fig:experiments}(a), the yellow root node at the top represents the source news, and the child nodes below are all reaction posts.
In rumor detection tasks, the problem is formulated as a binary classification task. The objective is to train a model using news instances labeled with ground-truth values $y \in \{\text{Rumor}, \text{Non-Rumor}\}$, enabling accurate prediction of labels for unseen test news items. Our work focuses on developing a model-agnostic framework for rumor detection, designed to seamlessly integrate with various graph learning algorithms and LLMs.
 
\section{Algorithm}
\label{headings}

\begin{figure*}[t]
  \centering
  \includegraphics[width=0.86\textwidth]{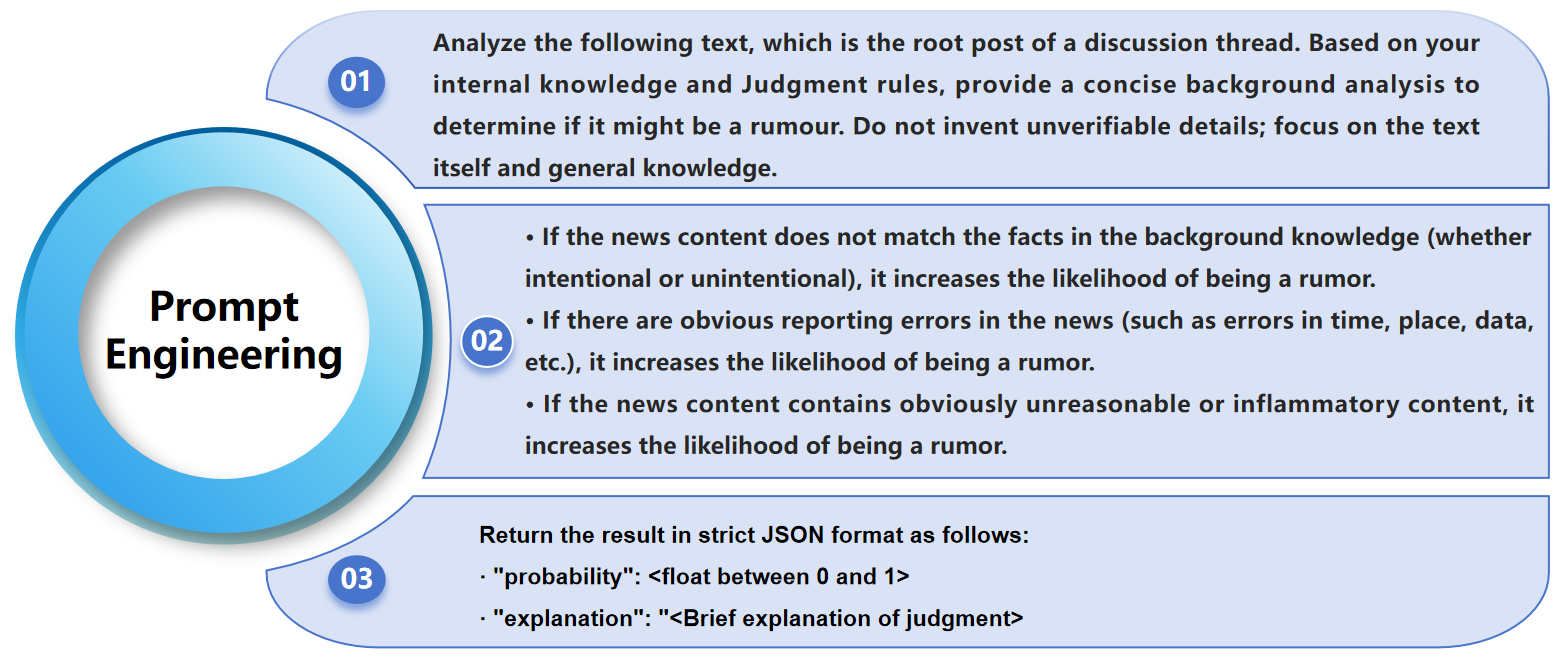}
  \caption{The LLM builds a shared event background from the root post, reuses it to score each subchain, and returns a probability with a brief explanation. Appendix H contains the full prompts.}
  \label{fig:prompt}
\end{figure*}

For each source news, a directed graph \(\mathbb{G} = (\mathbb{V}, \mathbb{E})\) is constructed:
\(\mathbb{V} = \{v_0, v_1, \dots, v_n\}\) is the set of nodes, where \(v_0\) is the source news (root node) and each child node \( v_i \in \mathbb{V} \setminus \{v_0\} \) represents a reply post. \( \mathbb{E} \) is the set of edges and the direction of the edge is the same as the direction of information propagation. The text associated with node \( v_i \) is denoted as \( \text{text}(v_i) \). Each node \(v_i\) is assigned a feature vector \(x_i\),   extracted using BERT from the final-layer [CLS] representation as follows:
$$ 
x_i = \text{BERT}(\text{text}(v_i)). 
$$

\textbf {Subchain Construction:} For each child node \( v_i \in \mathbb{V} \setminus \{v_0\} \), there exists a unique path (subchain) from the root node \( v_0 \) to \( v_i \), denoted as:
\[
\text{path}(v_0, v_i) = (v_0, v_{p_1}, v_{p_2}, \dots, v_{p_k}, v_i),
\]
where \( v_{p_j} \) are intermediate nodes on the path, and \( k \geq 0 \). The text information of the subchain is concatenated using a separator token [SEP]:
\[
\begin{aligned}
\text{info}(\text{path}(v_0, v_i)) = &\ \text{text}(v_0) \text{ [SEP] } \text{text}(v_{p_1})\cdots\\
&\ \cdots \text{ [SEP] } \text{text}(v_i).
\end{aligned}
\]

This chain-like information flow captures rich contextual information from the root to each child node. By leveraging this structure, our approach aims to enhance the capability of LLMs and graph learning algorithms to detect rumors effectively. As depicted in Figure~\ref{fig:experiments}(b), the red path from the root node to child node 2 represents the subchain associated with child node 2. By analogy, the entire graph contains as many subchains as there are child nodes, with each subchain corresponding to the unique path from the root node to a child node.

\begin{figure*}[t]
  \centering
  \includegraphics[width=0.90\textwidth]{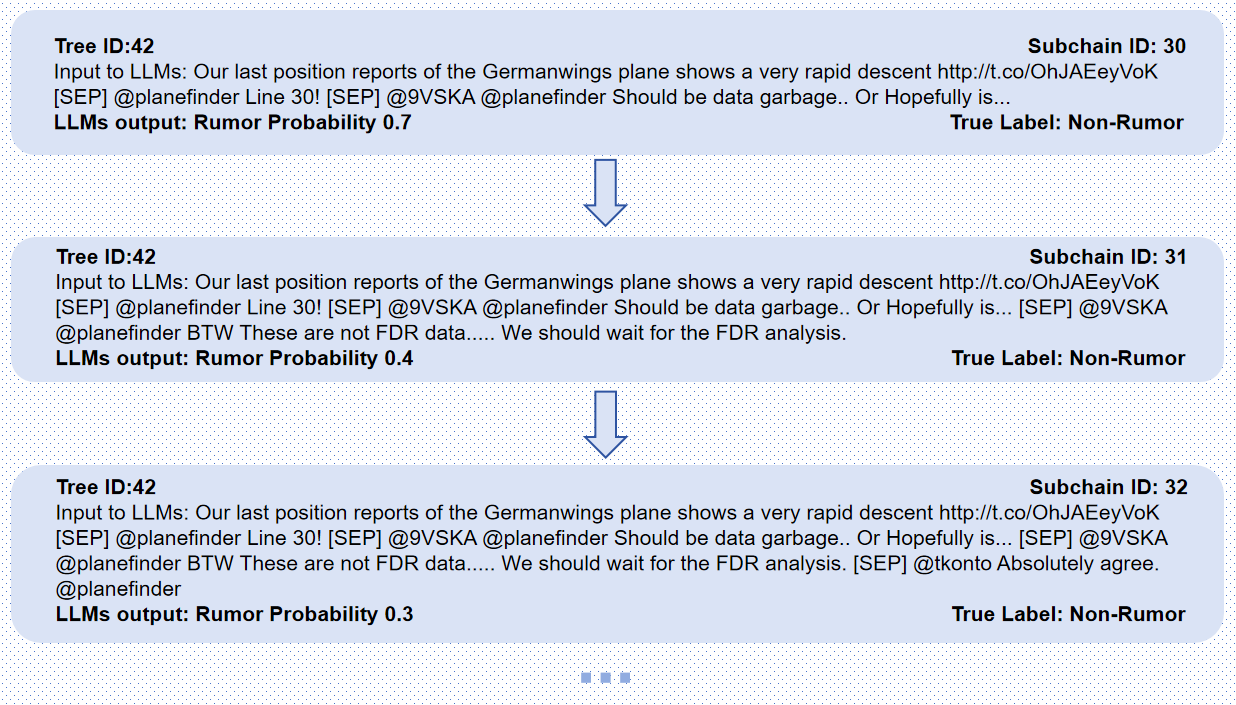}
  \caption{LLM rumor scores for three successive subchains of the Germanwings crash discussion. Each successive subchain adds one reply. The true label is Non-Rumor.}
  \label{fig:subchain-scoring}
\end{figure*}

\textbf{Virtual Node and Edge Augmentation in Graph Structures:}
To leverage the capabilities of LLMs and the graphical structure of information flows, we introduced a virtual node \( v^\ast \), labeled ``is Rumor",  which initially lacks feature information. Initial feature of \( v^\ast \) is set to the zero vector and its neighbors do not aggregate messages from \( v^\ast \). This means that even in a bidirectional GNN, a virtual node only has edges pointing to it.
We then use a LLM to process each subchain and outputs the probability that the source news is a rumor. For each child node \( v_i \), the LLM takes the concatenated text information of the subchain and outputs a probability:
\begin{eqnarray}\label{llmp0}
P_{\text{rumor}}(v_i) = \text{LLM}(\text{info}(\text{path}(v_0, v_i))),
\end{eqnarray} 
where \( P_{\text{rumor}}(v_i) \in [0, 1] \) represents the likelihood that the source news is a rumor based on the subchain from the root node to node \( v_i \). For each child node \( v_i \), if the LLM probability exceeds a predefined threshold \( \theta \):
\begin{eqnarray*}\label{llmp}
P_{\text{rumor}}(v_i) > \theta,
\end{eqnarray*} 
the directed edge \((v_i,v^\ast)\) is included in the threshold-selected set \(E_\theta\). Let \(\mathbb{E}'\) denote the final set of added edges after the minimum-edge rule below. The updated graph is \(\mathbb{G}' = (\mathbb{V} \cup \{v^\ast\}, \mathbb{E} \cup \mathbb{E}')\).
As shown in Figure~\ref{fig:experiments}(b), assume that the LLM assigns probabilities to the subchains of child nodes 1, 4, and 7 exceeding the threshold, edges are established between these nodes and the virtual node. Through these operations, we enhance the graph structure by introducing a virtual node and leveraging LLMs' reasoning capabilities to connection it to rumor propagation across the subchains. The rumor detection problem then reduces to predicting whether there is a link between the root node $v_0$ and the virtual node $v^*$.
Next we provide further details on how to retrieve the probability (\ref{llmp0}) from a LLM. 
It has been shown that carefully designed prompt engineering can enhance the ability of LLMs to identify rumors \citep{yan2024,shehata2024}. In our paper, we propose a specialized prompt engineering framework. 
Misinformation judgments can vary across LLMs and prompting strategies \citep{Huang2025unmasking}. In Step 1 of Figure~\ref{fig:prompt}, the prompt guides the LLM to analyze the root post using its internal knowledge and predefined criteria. The LLM compiles event background information into a shared knowledge base. In Step 2, the LLM uses the same context and criteria to evaluate every subchain in the tree.
As will be seen in Section 5.3, this prompting strategy successfully enhances the reliability of LLMs for rumor detection.
Figure~\ref{fig:subchain-scoring} presents an example illustrating how the LLM evaluates rumor probabilities across subchains in a propagation tree, using breaking news coverage of the Airbus A320 Germanwings crash. The true label of the root news is "Non-rumor". Subchains 31 and 32 each add one reply to the preceding subchain. We can see a clear downward trend in probability as the number of child nodes in the subchain increases. This sequential evaluation demonstrates the model's capability to process evolving information threads, dynamically adjusting probabilities based on accumulated context while maintaining consistency through the established knowledge base. The complete prompt templates are provided in Appendix H for full reproducibility.

We maintain a minimum number of virtual edges so that the virtual node can aggregate evidence even when few subchains score above the threshold. For child nodes \(v_1,\dots,v_n\), let \(p_i = P_{\text{rumor}}(v_i)\) and \(E_\theta = \{(v_i,v^\ast) \mid p_i > \theta\}\). The final set of added edges is
\[
\mathbb{E}' =
\begin{cases}
E_\theta, & \text{if } |E_\theta| \ge k, \\
\{(v_{\sigma(j)},v^\ast) \mid 1 \le j \le k\}, & \text{otherwise.}
\end{cases}
\]
Here, $ v^\ast $ denotes the virtual node, $ \sigma $ is the permutation sorting indices by $ p_{\sigma(1)} \geq \dots \geq p_{\sigma(n)} $, and $ k = \lceil \gamma n \rceil $ ensures at least a proportional subset of edges.  In Appendix D, we provide an ablation study  showing that  $\gamma = 0.2$ yields consistently stable performance, and that the root-virtual link prediction outperforms root-only classification under the same backbone. Appendix E further analyzes virtual-node connectivity and depth effects, while Appendix F reports subchain token statistics to characterize practical input sizes. Both provide further support for the robustness and promising performance of our method.

\begin{figure*}[!t]
  \centering
  \begin{tikzpicture}
    \node[anchor=south west,inner sep=0,outer sep=0] (original) at (0,0)
      {\includegraphics[width=0.88\textwidth]{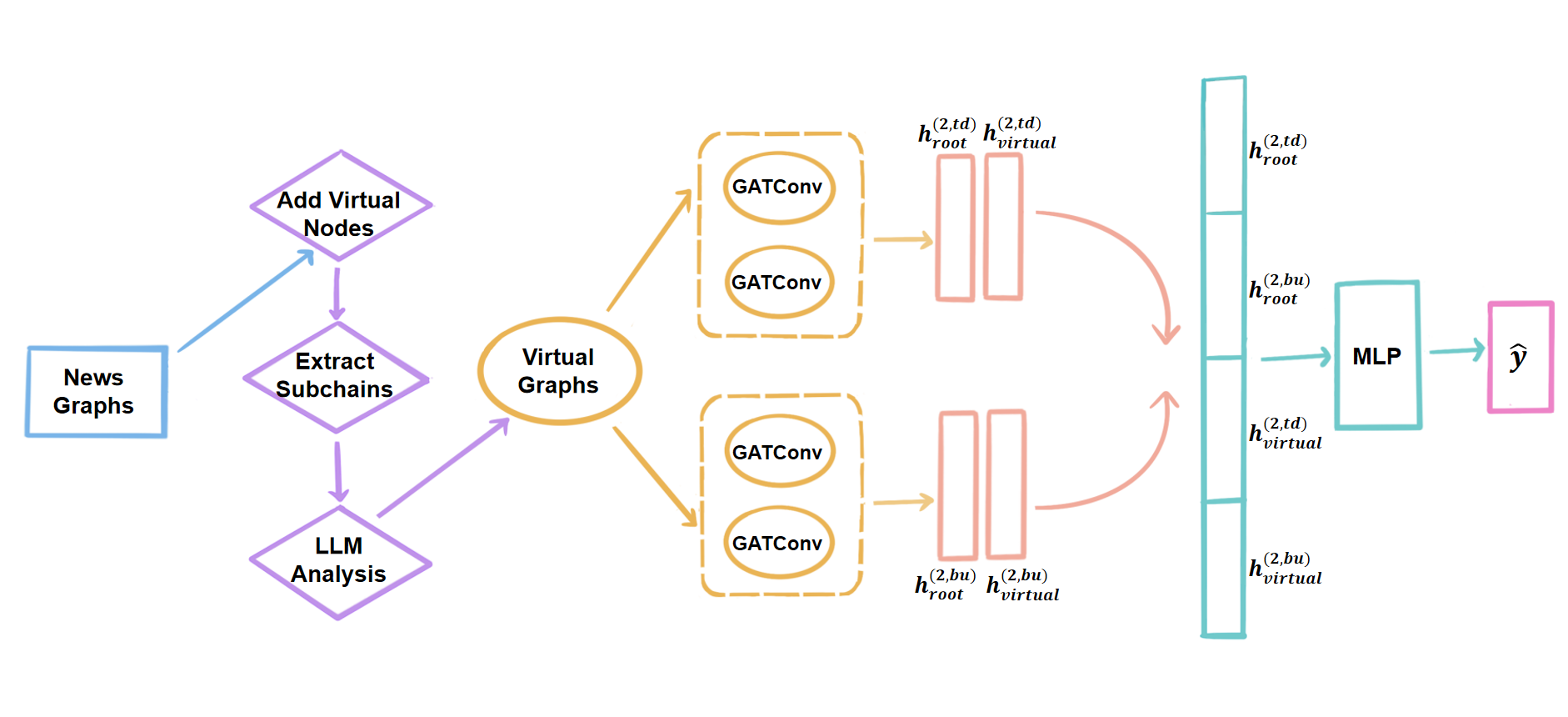}};
    \begin{scope}[x={(original.south east)},y={(original.north west)}]
      \node[font=\footnotesize\sffamily\bfseries,inner sep=0] at (0.779,0.943)
        {Concatenation};
    \end{scope}
  \end{tikzpicture}
  \caption{LLM-enhanced Bi-GAT. The root and virtual-node representations from both directions are concatenated and passed to the MLP to predict the root--virtual edge.}
  \label{fig:Algorithm}
\end{figure*}

We illustrate the framework using a bidirectional graph attention network (Bi-GAT) as the GNN backbone (Figure~\ref{fig:Algorithm}).
To implement Bi-GAT with the virtual node, the given graph \(\mathbb{G} = (\mathbb{V}, \mathbb{E})\) is extended to \( \tilde{\mathbb{G}} = (\tilde{\mathbb{V}}, \tilde{\mathbb{E}})\), with \( \tilde{\mathbb{V}} =  \mathbb{V} \cup \{v^\ast\}  \) and \( \tilde{\mathbb{E}}=   \{ (v_i, v_j), (v_j, v_i) : (v_i, v_j) \in \mathbb{E} \}  \cup\mathbb{E}'  \).

\paragraph{Graph Attention Convolution (GATConv)}
The GATConv function implements the graph attention mechanism proposed by \citet{velickovic2018graph}, which computes node representations by attending over neighboring nodes with learned attention coefficients. For a graph \( \tilde{\mathbb{G}} = (\tilde{\mathbb{V}}, \tilde{\mathbb{E}})\) with node features \(\textbf{x}_i \in \mathbb{R}^{F}\) for node \(v_i \in \tilde{\mathbb{V}}\) (node feature is set to be zero for $v^\ast$), the \texttt{GATConv} layer computes attention scores for its neighbors \(v_j \in \mathcal{N}(i)\), where \(\mathcal{N}(i) = \{v_j \mid (v_i, v_j) \in \tilde{\mathbb{E}}\}\) is the set of neighboring nodes. The attention score \(e_{ij}\) between nodes \(v_i\) and \(v_j\) is calculated as:
$$
e_{ij} = \text{LeakyReLU}\left(\mathbf{a}^T [\mathbf{W}_q \mathbf{x}_i \parallel \mathbf{W}_k \mathbf{x}_j]\right),
$$
where $\mathbf{W}_q$, $\mathbf{W}_k$, $\mathbf{W}_v\in \mathbb{R}^{F' \times F}$ are weight matrices that transform the input features to the hidden dimension. \(\mathbf{a} \in \mathbb{R}^{2 F'}\) is the attention parameter vector. \([\mathbf{W}_q \mathbf{x}_i \parallel \mathbf{W}_k \mathbf{x}_j] \in \mathbb{R}^{2 F'}\) is the concatenation of the transformed features of nodes \(v_i\) and \(v_j\). Then the attention coefficients \(\alpha_{ij}\) are normalized across neighbors using the softmax function and use the attention coefficient to weight the neighbor value vector to update the feature of node $i$:
\[
\alpha_{ij} = \frac{\exp(e_{ij})}{\sum_k \exp(e_{ik})},\quad
\mathbf{h}_i = \sum_j \alpha_{ij} \mathbf{W}_v x_j,
\]
where $k,j \in \mathcal{N}(i)$. To stabilize and enhance the attention mechanism, multiple attention heads are employed. Each head \(h = 1, \dots, H\)  has its own weight matrix \(\mathbf{W}_v^{(h)} \in \mathbb{R}^{F' \times F}\) and attention vector \(\mathbf{a}^{(h)} \in \mathbb{R}^{2 F'}\). The output of each head is:
\begin{align*}
\mathbf{h}_i^{(h)} &= \sum_{j \in \mathcal{N}(i)} \alpha_{ij}^{(h)} \mathbf{W}_v^{(h)} \mathbf{x}_j, \\
\mathbf{h}_i &= \parallel_{h=1}^H \mathbf{h}_i^{(h)} \in \mathbb{R}^{HF'},
\end{align*}
where \(\alpha_{ij}^{(h)}\) is the attention coefficient for head \(h\). $\textbf{h}_i\in \mathbb{R}^{HF'}$ is the final output for node \(v_i\) concatenating the head outputs. 

\paragraph{Top-Down GAT (TD-GAT)} Simulate the information transmission from the ``high-level" nodes to the ``low-level" nodes of the graph, which is suitable for capturing causal relationships.
\begin{itemize}
    \item \textbf{First GATConv Layer}:
\end{itemize}
$$  
     \mathbf{h}_i^{(1, \text{td})} = \text{ReLU}\left(\parallel_{h=1}^H \sum_{j \in \mathcal{N}_{out}(i)} \alpha_{ij}^{(h)} \mathbf{W}_{\text{td1}}^{(h)} \mathbf{x}_j\right),
$$
where $\mathcal{N}_{out}(i)$ is the out-neighbors of node i and $\mathbf{h}_i^{(1, \text{td})}\in\mathbb{R}^{HF'}$.
\begin{itemize}
     \item \textbf{Second GATConv Layer}:
\end{itemize}
$$
    \mathbf{h}_i^{(2, \text{td})} = \frac{1}{H}\sum_{h=1}^H\sum_{j \in \mathcal{N}_{out}(i)} \alpha_{ij}\mathbf{W}_{\text{td2}} [\mathbf{h}_j^{(1, \text{td})} \parallel \mathbf{x}_{\text{r}}] ,
$$
where $x_{r}$ is the feature of root node and $\mathbf{h}_i^{(2, \text{td})}\in\mathbb{R}^{F'}$.

\paragraph{Bottom-Up GAT (BU-GAT)} Simulate the aggregation of features from low-level nodes to high-level nodes of the graph.
\begin{itemize}
        \item \textbf{First GATConv Layer}:
\end{itemize}
        $$
        \mathbf{h}_i^{(1, \text{bu})} = \text{ReLU}\left(\parallel_{h=1}^H \sum_{j \in \mathcal{N}_{in}(i)} \alpha_{ij}^{(h)} \mathbf{W}_{\text{bu1}}^{(h)} \mathbf{x}_j\right),
        $$
        where \(\mathcal{N}_{in}(i)\) is the in-neighbors of node i.
\begin{itemize}
        \item \textbf{Second GATConv Layer}:
\end{itemize}
        $$
        \mathbf{h}_i^{(2, \text{bu})} = \frac{1}{H}\sum_{h=1}^H\sum_{j \in \mathcal{N}_{in}(i)} \alpha_{ij} \mathbf{W}_{\text{bu2}} [\mathbf{h}_j^{(1, \text{bu})} \parallel \mathbf{x}_{\text{r}}].
        $$

\paragraph{Feature Fusion and Edge Classification}
\begin{itemize}
    \item \textbf{Feature Extraction}: Extract \(\mathbf{h}_{\text{root}}^{(2, \text{td})}\), \(\mathbf{h}_{\text{virtual}}^{(2, \text{td})}\), \(\mathbf{h}_{\text{root}}^{(2, \text{bu})}\), and \(\mathbf{h}_{\text{virtual}}^{(2, \text{bu})}\).
    \item \textbf{Fusion}: Concatenate into 
    \(\mathbf{c} = [\mathbf{h}_{\text{root}}^{(2, \text{td})}, \mathbf{h}_{\text{root}}^{(2, \text{bu})}, \mathbf{h}_{\text{virtual}}^{(2, \text{td})}, \mathbf{h}_{\text{virtual}}^{(2, \text{bu})}] \in \mathbb{R}^{4F'}\).
    \item \textbf{Classification}:
    $
    \mathbf{z} = \mathbf{W}_{\text{edge}} \mathbf{c} + \mathbf{b}_{\text{edge}},$  
    $
    \hat{y}_{\text{edge}} = \sigma(\textbf{z}),
    $
    where \(\mathbf{W}_{\text{edge}} \in \mathbb{R}^{4F'}\), \(\mathbf{b}_{\text{edge}} \in \mathbb{R}\), and \(\sigma\) is the sigmoid function.
\end{itemize}

\paragraph{Model and Training} We use the binary cross-entropy function. For a batch size of N, the loss function is defined as
\begin{align*}
\textbf{Loss} = -\frac{1}{N}\sum_{i=1}^N\omega_i \big[  y_i&\log(\sigma(z_i)) \\
+ & (1-y_i)\log(1-\sigma(z_i)) \big]
\end{align*}
where $z_i$ is the positive class logit output by the model (i.e. the logit score of the virtual node and the root node with an edge). $y_i$ is the true label of the source news, and $\omega_i$ is the weight determined by $y_i$'s category and pos\_weight. 
pos\_weight is set to the negative-to-positive class ratio, which increases the loss weight for positive (minority) samples to prioritize their correct prediction and counteract class imbalance. Parameters are trained using the Adam optimizer with backpropagation to optimize all components of the Bi-GAT model. Each backbone and its variants share a separate 7:1:2 train/validation/test split. We select $\theta$ only on validation data. For PHEME, we apply overall stratified sampling on the full dataset (not per event), then pool validation instances from all five events to choose one shared threshold via Youden’s (J). For Weibo, which has no event partition, we similarly use a global split and one validation-selected threshold. The hyperparameters were set as follows: learning rate = 0.00005, weight decay = 1e-3, dropout rate = 0.3, and a maximum of 150 training epochs. An early stopping mechanism was used; during training, the F1 score on the validation set is continuously monitored, and training is automatically stopped if the score failed to exceed the historical best value for 20 consecutive epochs, thus preventing model overfitting.

\section{Experiments}
\subsection{Datasets and LLMs}
We evaluate on five news events from PHEME (Charlie Hebdo, Ferguson, Germanwings Crash, Ottawa Shooting, and Sydney Siege) and on the Weibo dataset \citep{zubiaga2016pheme,ma2016}. Table~1 summarizes both public datasets. We remove graphs without replies and apply the same preprocessing to all baselines. We use DeepSeek-V3 as the base LLM and report additional Qwen-Plus results in Appendices B and C.
\begin{table}[ht]
\centering
\footnotesize 
\scalebox{0.90}{ 
\begin{tabular}{l c c c}
\toprule
\textbf{News Event} & \textbf{Non-Rumor} & \textbf{Rumor} & \textbf{Total} \\ 
\midrule
Charlie Hebdo      & 1621 & 458  & 2079 \\ 
Ferguson           & 859  & 284  & 1143 \\ 
Germanwings Crash  & 231  & 238  & 469  \\ 
Ottawa Shooting    & 420  & 470  & 890  \\ 
Sydney Siege       & 699  & 522  & 1221 \\ 
Weibo              & 2351 & 2313 & 4664 \\
\bottomrule
\end{tabular}
}
\caption{Statistics of PHEME and Weibo Datasets.}
\end{table}

\subsection{Evaluation metrics}
Table 2 presents the fundamental metrics for assessing classification model performance: Accuracy, Precision, Recall, and F1 Score.

\begin{table}[ht]
\centering
\footnotesize 
\scalebox{0.8}{ 
\begin{tabular}{l c l l}
\toprule
\textbf{Metric} & \textbf{Formula} & \textbf{Term}  \\ 
\midrule
Accuracy & $\dfrac{\text{TP} + \text{TN}}{\text{TP} + \text{TN} + \text{FP} + \text{FN}}$ & TP  \\ 
Precision & $\dfrac{\text{TP}}{\text{TP} + \text{FP}}$ & TN \\ 
Recall & $\dfrac{\text{TP}}{\text{TP} + \text{FN}}$ & FP  \\ 
F1 Score & $2 \cdot \dfrac{\text{Precision} \cdot \text{Recall}}{\text{Precision} + \text{Recall}}$ & FN  \\ 
\bottomrule
\end{tabular}
}
\caption{Classification Metrics, Terms, and Formulas.}
\end{table}

We name our proposed framework LLM-VN and compare it with existing methods as comparative baselines:
Bi-GCN \citep{bian2020} captures bidirectional dependencies in trees;
RvNN \citep{ma2018rvnn} recursively aggregates post representations along propagation trees;
GAT \citep{velickovic2018graph} uses attention to weigh neighbor importance;
HD-TRANs \citep{ma2020debunking} uses tree-structured attention to aggregate signals across replies;
LINE \citep{tang2015line} provides large-scale static network embeddings;
DDGCN \citep{korban2020ddgcn} is a dynamic directed GCN originally developed for action recognition;
GraphSAGE \citep{hamilton2017inductive} samples and aggregates local features inductively;
GSMA \citep{ma2024rumor} enhances GraphSAGE with attention and position encoding for rumor detection.
SePro\citep{zeng2025} refines social contexts via semantic-propagation collaboration and Chain-of-Clue prompting, improving LLMs' performance.

The `LLM-VN' enhanced version of the above baselines are denoted  as `{\bf LLM-VN+Baseline}'. We also compare our approach to simple model average where the final probability of rumors is a weighted average of the probability from LLM with well-designed prompt and the probability from the baseline method. The optimal weight is tuned via line search. We denote this as `{\bf LLM+Baseline}'.
 
\subsection{Results and analysis}
As shown in Table 3, on the PHEME dataset, LLMs perform poorly in rumor identification without carefully designed prompts. Acc. (R) measures the proportion of correctly identified ``rumor'' instances among all true rumors, while Acc. (N) does the same for ``non-rumor'' instances.

\begin{table}[ht]
\centering
\scalebox{0.80}{ 
\begin{tabular}{lcc}
\toprule
News Event & Acc. (R) & Acc. (N) \\
\midrule
Charlie Hebdo & 15.54\% & 83.20\% \\
Ferguson & 41.55\% & 89.76\% \\
Germanwings crash & 23.11\% & 91.77\% \\
Ottawa shooting & 16.81\% & 87.62\% \\
Sydney Siege & 33.14\% & 78.25\% \\
\bottomrule
\end{tabular}
}
\caption{Performance of LLMs under basic prompts.}
\end{table}

 The performance of our proposed framework `LLM-VN' combined with Bi-GAT on the PHEME dataset are presented in Table 4:

\begin{table}[ht]
\centering
\scalebox{0.75}{  
\begin{tabular}{l *{5}{c}}
\toprule
\textbf{News Event} & {\textbf{Acc.}} & {\textbf{Pre.}} & {\textbf{Rec.}} & {\textbf{F1}} & {\textbf{AUC}} \\ 
\midrule
Charlie Hebdo      & 0.923 & 0.943 & 0.967 & 0.955 & 0.962 \\ 
Ferguson           & 0.860 & 0.818 & 0.892 & 0.853 & 0.903 \\ 
Germanwings Crash  & 0.894 & 0.914 & 0.961 & 0.937 & 0.827 \\ 
Ottawa Shooting    & 0.876 & 0.883 & 0.987 & 0.932 & 0.846 \\ 
Sydney Siege       & 0.829 & 0.866 & 0.905 & 0.885 & 0.858 \\ 
\bottomrule
\end{tabular}
}
\caption{Performance of LLM-VN+Bi-GAT.}
\end{table}

\begin{table*}[ht]
\centering
\scalebox{0.85}{
\small
\setlength{\tabcolsep}{3pt}
\begin{tabular}{l|ccccc|ccccc}
\toprule
& \multicolumn{5}{c|}{\textbf{PHEME Dataset}} & \multicolumn{5}{c}{\textbf{Weibo Dataset}} \\
\cmidrule(lr){2-6} \cmidrule(lr){7-11}
\textbf{Method} & \textbf{Acc} & \textbf{Prec} & \textbf{Rec} & \textbf{F1} & \textbf{Class} & \textbf{Acc} & \textbf{Prec} & \textbf{Rec} & \textbf{F1} & \textbf{Class} \\
\midrule
Bi-GCN & 0.824 & 0.752/0.861 & 0.734/0.872 & 0.742/0.866 & R/N & 0.963 & 0.951/0.970 & 0.946/0.973 & 0.948/0.971 & R/N \\
LLM+Bi-GCN & 0.830 & 0.758/0.866 & 0.744/0.874 & 0.751/0.870 & R/N & 0.971 & 0.961/0.976 & 0.958/0.978 & 0.959/0.977 & R/N \\
LLM-VN+Bi-GCN & \textbf{0.842} & 0.781/0.872 & 0.752/0.889 & 0.766/0.880 & R/N & \textbf{0.988} & 0.982/0.991 & 0.985/0.990 & 0.983/0.991 & R/N \\
\addlinespace
RvNN & 0.763 & 0.730/0.777 & 0.589/0.868 & 0.652/0.820 & R/N & 0.771 & 0.727/0.797 & 0.681/0.830 & 0.703/0.813 & R/N \\
LLM+RvNN & 0.783 & 0.698/0.842 & 0.754/0.802 & 0.725/0.821 & R/N & 0.775 & 0.742/0.793 & 0.668/0.846 & 0.703/0.819 & R/N \\
LLM-VN+RvNN & \textbf{0.813} & 0.782/0.830 & 0.704/0.880 & 0.740/0.854 & R/N & \textbf{0.780} & 0.697/0.850 & 0.796/0.770 & 0.743/0.808 & R/N \\
\addlinespace
Graphsage & 0.842 & 0.787/0.880 & 0.820/0.857 & 0.803/0.868 & R/N & 0.963 & 0.948/0.972 & 0.944/0.974 & 0.946/0.973 & R/N \\
GSMA & 0.848 & 0.798/0.882 & 0.823/0.864 & 0.810/0.873 & R/N & 0.974 & 0.967/0.976 & 0.953/0.984 & 0.960/0.980 & R/N \\
LLM-VN+Graphsage & \textbf{0.866} & 0.831/0.887 & 0.825/0.891 & 0.828/0.889 & R/N & \textbf{0.981} & 0.991/0.975 & 0.950/0.996 & 0.970/0.985 & R/N \\
\addlinespace
GAT & 0.811 & 0.737/0.864 & 0.796/0.820 & 0.765/0.841 & R/N & 0.947 & 0.939/0.953 & 0.936/0.954 & 0.938/0.954 & R/N \\
LLM+GAT & 0.823 & 0.761/0.866 & 0.796/0.841 & 0.778/0.853 & R/N & 0.951 & 0.944/0.956 & 0.940/0.958 & 0.942/0.957 & R/N \\
SePro & 0.831 & 0.793/0.852 & 0.762/0.874 & 0.777/0.863 & R/N & 0.950 & 0.938/0.960 & 0.948/0.953 & 0.943/0.956 & R/N \\
LLM-VN+GAT & \textbf{0.847} & 0.814/0.868 & 0.788/0.886 & 0.800/0.877 & R/N & \textbf{0.982} & 0.981/0.984 & 0.978/0.985 & 0.980/0.985 & R/N \\
\addlinespace
HD-TRANs & 0.766 & 0.751/0.779 & 0.755/0.775 & 0.753/0.777 & R/N & 0.974 & 0.954/0.981 & 0.946/0.984 & 0.950/0.983 & R/N \\
LLM+HD-TRANs & 0.779 & 0.767/0.789 & 0.766/0.789 & 0.767/0.789 & R/N & 0.980 & 0.965/0.984 & 0.953/0.988 & 0.959/0.986 & R/N \\
LLM-VN+HD-TRANs & \textbf{0.796} & 0.779/0.811 & 0.793/0.797 & 0.786/0.804 & R/N & \textbf{0.991} & 0.988/0.991 & 0.974/0.996 & 0.981/0.994 & R/N \\
\addlinespace
LINE & 0.744 & 0.732/0.754 & 0.730/0.755 & 0.731/0.754 & R/N & 0.790 & 0.774/0.802 & 0.771/0.806 & 0.772/0.804 & R/N \\
LLM+LINE & 0.756 & 0.742/0.768 & 0.750/0.760 & 0.746/0.764 & R/N & 0.803 & 0.794/0.809 & 0.775/0.826 & 0.784/0.818 & R/N \\
LLM-VN+LINE & \textbf{0.786} & 0.773/0.798 & 0.781/0.789 & 0.777/0.794 & R/N & \textbf{0.811} & 0.793/0.828 & 0.804/0.818 & 0.798/0.823 & R/N \\
\addlinespace
DDGCN & 0.855 & 0.877/0.836 & 0.809/0.896 & 0.842/0.865 & R/N & 0.951 & 0.947/0.953 & 0.933/0.963 & 0.940/0.958 & R/N \\
LLM+DDGCN & 0.860 & 0.855/0.865 & 0.852/0.866 & 0.854/0.865 & R/N & 0.954 & 0.943/0.963 & 0.949/0.958 & 0.946/0.961 & R/N \\
LLM-VN+DDGCN & \textbf{0.876} & 0.876/0.875 & 0.862/0.887 & 0.869/0.881 & R/N & \textbf{0.984} & 0.983/0.985 & 0.978/0.987 & 0.981/0.986 & R/N \\
\bottomrule
\end{tabular}
}
\caption{Performance Comparison on PHEME and Weibo Datasets.}
\end{table*}

From Table 4, the model achieves strong performance on most news events, with accuracy often exceeding 85\%, along with high precision, recall, and F1 scores.
To better illustrate the model-agnostic nature of our approach and demonstrate the added benefits of the virtual node beyond  prompt engineering, we present in 
Table 5 a comprehensive comparison of graph learning methods for rumor detection on the Weibo and PHEME datasets. It evaluates baseline models in three setups: the original baseline, LLM+Baseline, and LLM-VN+Baseline. To ensure a fair comparison, our proposed prompting strategy (Appendix H) is applied to all evaluated baselines. 
We follow the standard benchmark setting on PHEME and Weibo, rather than a strictly time-grounded early detection protocol. In our framework, pretrained LLM knowledge serves only as an auxiliary signal for enhancing GNN-based rumor detection, while the final prediction is still made by the downstream graph model. 
From table 5, we can see that the LLM-VN-enhanced methods demonstrate significant advantages on both the PHEME and Weibo datasets. Compared to the original baseline method and the model average approaches (LLM+Baseline), our framework shows substantial improvements in key metrics such as accuracy, precision, recall, and F1 score.  Especially on the PHEME dataset, methods like LLM-VN+DDGCN  exhibit particularly outstanding performance in accuracy and F1 score. On the Weibo dataset, methods such as LLM-VN+Bi-GCN and LLM-VN+GAT also outperform the baseline, particularly in terms of precision and recall for the non-rumor  category, demonstrating higher classification stability. In contrast, the improvement of LLM+Baseline is less robust than its VN-enhanced version, especially in cases of complex propagation structures or imbalanced data distributions. On the Weibo dataset, methods outperform PHEME due to Weibo's larger, deeper, and richer propagation trees. Overall, the LLM-VN framework enhances generalization and classification accuracy by integrating LLMs' semantic understanding with graph models' structured processing, particularly on complex social media propagation patterns.

\subsection{Subchain Sampling and Threshold Sensitivity}

On PHEME, LLM-VN+Bi-GAT outperforms Bi-GAT in accuracy and F1 on all five events when the LLM scores 60\% of the root-to-reply subchains (Table~\ref{tab:subchain-sampling}). Mean F1 across events rises from 0.828 for Bi-GAT to 0.890 with 60\% scoring and reaches 0.912 with full scoring.

\begin{table}[!ht]
\centering
\footnotesize
\setlength{\tabcolsep}{2.5pt}
\begin{tabular}{@{}lccc@{}}
\toprule
& \multicolumn{2}{c}{\textbf{LLM-VN+Bi-GAT}} & \\
\cmidrule(lr){2-3}
\textbf{Event} & \textbf{100\%} & \textbf{60\%} & \textbf{Bi-GAT} \\
\midrule
\multicolumn{4}{@{}l}{\textit{Accuracy}} \\
Charlie Hebdo     & 0.923 & $0.901 \pm 0.013$ & $0.868 \pm 0.026$ \\
Ferguson          & 0.860 & $0.832 \pm 0.019$ & $0.785 \pm 0.033$ \\
Germanwings Crash & 0.894 & $0.872 \pm 0.015$ & $0.799 \pm 0.028$ \\
Ottawa Shooting   & 0.876 & $0.853 \pm 0.017$ & $0.811 \pm 0.031$ \\
Sydney Siege      & 0.829 & $0.804 \pm 0.020$ & $0.744 \pm 0.037$ \\
\addlinespace
\multicolumn{4}{@{}l}{\textit{F1}} \\
Charlie Hebdo     & 0.955 & $0.936 \pm 0.011$ & $0.895 \pm 0.024$ \\
Ferguson          & 0.853 & $0.823 \pm 0.021$ & $0.763 \pm 0.038$ \\
Germanwings Crash & 0.937 & $0.917 \pm 0.013$ & $0.837 \pm 0.027$ \\
Ottawa Shooting   & 0.932 & $0.911 \pm 0.014$ & $0.852 \pm 0.030$ \\
Sydney Siege      & 0.885 & $0.863 \pm 0.018$ & $0.795 \pm 0.034$ \\
\bottomrule
\end{tabular}
\caption{Accuracy and F1 on PHEME by subchain scoring coverage.}
\label{tab:subchain-sampling}
\end{table}

With DeepSeek-V3 and Bi-GAT, we vary $\theta$ around $\theta^\star$, selected on validation data using Youden's $J$. We fix the LLM scores, data split, backbone, $\gamma$, and training protocol. Metrics are first computed per event, then averaged. Among the five settings, $\theta^\star$ gives the highest accuracy, precision, F1, and AUC (Table~\ref{tab:threshold-sensitivity}).

\begin{table}[H]
\centering
\footnotesize
\setlength{\tabcolsep}{3pt}
\begin{tabular}{@{}lrrrrr@{}}
\toprule
$\theta$ & Acc. & Prec. & Rec. & F1 & AUC \\
\midrule
$\theta^\star - 0.2$ & 0.8082 & 0.6846 & 0.8182 & 0.7454 & 0.8879 \\
$\theta^\star - 0.1$ & 0.8357 & 0.7327 & 0.8209 & 0.7743 & 0.9169 \\
$\theta^\star$       & \textbf{0.8469} & \textbf{0.7694} & 0.8121 & \textbf{0.7902} & \textbf{0.9288} \\
$\theta^\star + 0.1$ & 0.8137 & 0.6957 & 0.8128 & 0.7497 & 0.8961 \\
$\theta^\star + 0.2$ & 0.8045 & 0.6739 & \textbf{0.8342} & 0.7455 & 0.8895 \\
\bottomrule
\end{tabular}
\caption{Test performance at the validation-selected threshold $\theta^\star$ and four nearby values.}
\label{tab:threshold-sensitivity}
\end{table}

\subsection{Computational Cost}

LLM scoring runs once during offline graph augmentation. The API calls for individual subchains are independent and can run in parallel. Table~\ref{tab:computational-cost} reports the measured wall-clock time and API cost for PHEME and Weibo with 100 concurrent requests on a single workstation. Appendix A gives the measurement setup and per-item timings.

\begin{table}[H]
\centering
\small
\begin{tabular}{lcc}
\toprule
\textbf{Metric} & \textbf{PHEME} & \textbf{Weibo} \\
\midrule
Trees processed & 5,447 & 3,805 \\
Posts processed & 96,344 & 3,804,357 \\
Wall-clock time & 40--50 min & 35--40 h \\
API cost (USD) & 5--10 & 80--100 \\
\bottomrule
\end{tabular}
\caption{Runtime and API cost of offline graph augmentation with 100 concurrent requests.}
\label{tab:computational-cost}
\end{table}

\section{Conclusions}
We propose a general framework to enhance the performance of graph learning methods for rumor prediction by leveraging LLMs. By employing LLMs to analyze subchains and assign rumor probabilities, we augment the graph with a virtual ``is Rumor" node. Intuitively, when LLMs predict high rumor probabilities across many subchains, the virtual node shares more neighbors with the root node, leading to convergent embeddings during GNN propagation. This enhances the likelihood of link prediction between the root and virtual nodes, effectively classifying the source news as a rumor.  A key strength of our design is its robustness to imperfect LLM signals: the mitigation mechanism reduces overreliance on noisy or weak evidence, even when the LLM covers only a subset of the subchains, it can still effectively improve the final prediction performance. 
Notably, our method is a modular framework that leverages LLMs to retrieve and reason over subchain evidence, injecting this signal into a propagation graph for GNN-based prediction. The model-agnostic design allows it to be paired with different GNN backbones and LLMs.

\section*{Limitations}
A limitation of our approach emerges when handling extremely large-scale graphs. The LLM API latency can prevent real-time responses for very large graphs (Table~\ref{tab:computational-cost}).


\section*{Acknowledgments}
We used AI-assisted tools to assist with language polishing and manuscript revision, the preliminary literature search, and, in the algorithm section, to provide reference probability estimates for rumor judgment based on our prompts. These tools were used only as auxiliary aids, the authors designed the study, interpreted the results, made all final judgments and decisions, and reviewed and verified all AI-assisted content. The authors are solely responsible for the content of this manuscript.




\section*{A$\quad$ Runtime Measurement Setup}
We measured the timings in Table~\ref{tab:computational-cost} with DeepSeek-V3 or Qwen-Plus API calls on a workstation with at least 16 CPU cores, 32 GB RAM, and a stable 100 Mbps network connection. With single-threaded API calls, both services processed more than 1,200 nodes per hour. Typical processing times per news item were about 10 seconds on PHEME and 30--60 seconds on Weibo.

\section*{B$\quad$ LLMs Bias}
To further demonstrate the biases inherent in LLMs, we evaluated their performance on the PHEME dataset using Qwen-Plus. In Table~\ref{tab:qwen-basic-prompts}, our analysis reveals that Qwen exhibits a markedly different bias compared to DeepSeek. This divergence primarily stems from their distinct training data and reinforcement learning strategies. Specifically, Qwen demonstrates a more subjective and aggressive stance, it tends to err on the side of caution by prioritizing the avoidance of false negatives, even at the cost of increasing false positives. In contrast, DeepSeek adopts a considerably more conservative and neutral approach, leaning towards minimizing false positives rather than risking false negatives.
\begin{table}[ht]
\centering
\scalebox{0.80}{ 
\begin{tabular}{lcc}
\toprule
News Event & Acc. (R) & Acc. (N) \\
\midrule
Charlie Hebdo & 92.3\% & 21.8\% \\
Ferguson &  88.7\% & 34.5\% \\
Germanwings crash &  94.1\% & 19.2\% \\
Ottawa shooting & 90.6\% & 26.7\% \\
Sydney Siege & 86.9\% & 31.4\% \\
\bottomrule
\end{tabular}
}
\caption{Performance of Qwen-Plus under basic prompts on PHEME.}
\label{tab:qwen-basic-prompts}
\end{table}

\section*{C$\quad$ Model-agnostic Claim}
We have conducted additional experiments using Qwen-Plus on the PHEME dataset to strengthen our model-agnostic claim. As shown in Table~\ref{tab:pheme_results}, our framework yields consistent improvements across baselines, with gains in accuracy and other metrics, closely mirroring the results obtained with DeepSeek-V3. We use the default temperature for both LLMs without validation tuning: 1.0 for DeepSeek and 0.7 for Qwen-Plus. In practice, the default setting is sufficient and avoids introducing another decoding hyperparameter. These results confirm the framework’s robustness across different LLM families.

\begin{table}[ht]
\centering
\scalebox{0.55}{  
\begin{tabular}{lccccc}
\toprule
Method & Acc & Prec & Rec & F1 & Class \\
\midrule
Bi-GCN & 0.824 & 0.752/0.861 & 0.734/0.872 & 0.742/0.866 & R/N \\
\textbf{LLM-VN+Bi-GCN} & \textbf{0.833} & 0.752/0.880 & 0.778/0.858 & 0.765/0.864 & R/N \\
\addlinespace
RvNN & 0.763 & 0.730/0.777 & 0.589/0.868 & 0.652/0.820 & R/N \\
\textbf{LLM-VN+RvNN} & \textbf{0.786} & 0.793/0.835 & 0.770/0.896 & 0.781/0.842 & R/N \\
\addlinespace
Graphsage & 0.842 & 0.787/0.880 & 0.820/0.857 & 0.803/0.868 & R/N \\
GSMA & 0.848 & 0.798/0.882 & 0.823/0.864 & 0.810/0.873 & R/N \\
\textbf{LLM-VN+Graphsage} & \textbf{0.852} & 0.803/0.882 & 0.821/0.871 & 0.811/0.876 & R/N \\
\addlinespace
GAT & 0.811 & 0.737/0.864 & 0.796/0.820 & 0.765/0.841& R/N \\
\textbf{LLM-VN+GAT} & \textbf{0.823} & 0.730/0.880 & 0.780/0.850 & 0.750/0.860 & R/N \\
\addlinespace
HD-TRANs & 0.766 & 0.751/0.779 & 0.755/0.775 & 0.753/0.777 & R/N \\
\textbf{LLM-VN+HD-TRANs} & \textbf{0.779} & 0.712/0.808 & 0.729/0.790 & 0.719/0.796 & R/N \\
\addlinespace
LINE & 0.744 & 0.732/0.754 & 0.730/0.755 & 0.731/0.754 & R/N \\
\textbf{LLM-VN+LINE} & \textbf{0.755} & 0.712/0.782 & 0.731/0.769 & 0.720/0.776 & R/N \\
\addlinespace
DDGCN & 0.855 & 0.877/0.836 & 0.809/0.896 & 0.842/0.865 & R/N \\
\textbf{LLM-VN+DDGCN} & \textbf{0.865} & 0.811/0.890 & 0.832/0.877 & 0.820/0.881 & R/N \\
\bottomrule
\end{tabular}%
}
\caption{Performance Comparison on PHEME Dataset via Qwen-Plus.}
\label{tab:pheme_results}
\end{table}
\section*{D$\quad$ Ablation Study}
We conducted an ablation study of $\gamma$ using Qwen-Plus + Bi-GAT on the PHEME dataset. The results in Table~\ref{tab:gamma_ablation} show that $\gamma=0.2$ consistently yields the best performance, striking an optimal balance between incorporating global context and preserving local structural signals.

\begin{table}[ht]
\centering
\scalebox{0.75}{  
\begin{tabular}{crrrrr}
\toprule
$\gamma$ (\%) & Acc. & Rec. & F1 & Prec. & AUC \\
\midrule
10 & 0.8138 & 0.7727 & 0.7401 & 0.7101 & 0.8767 \\
15 & 0.8083 & 0.8021 & 0.7417 & 0.6897 & 0.8780 \\
\textbf{20} & \textbf{0.8229} & 0.7754 & \textbf{0.7503} & \textbf{0.7268} & \textbf{0.8798} \\
25 & 0.8128 & 0.7888 & 0.7431 & 0.7024 & 0.8763 \\
30 & 0.8018 & \textbf{0.8155} & 0.7385 & 0.6748 & 0.8744 \\
\bottomrule
\end{tabular}
}
\caption{Ablation study on the virtual node mixing coefficient  $\gamma$.}
\label{tab:gamma_ablation}
\end{table}

We further include a control experiment comparing two prediction formulations under the same backbone and training protocol. 
Specifically, we contrast (i) root-only classification, which predicts the rumor label directly from the root representation, with (ii) root-virtual link prediction, which predicts whether the root node is connected to the virtual node after virtual-edge augmentation. 
Since the encoder and data processing are kept identical, this comparison directly tests whether the link-prediction formulation is more effective than classifying from the root alone.\\

\begin{table}[t]
\centering
\scriptsize
\setlength{\tabcolsep}{5pt}
\resizebox{0.95\linewidth}{!}{%
\begin{tabular}{lcc}
\hline
\textbf{Event} & \textbf{Root-only Acc.} & \textbf{Link Pred. Acc.} \\
\hline
Charlie Hebdo & 0.905 & 0.923 \\
Ferguson & 0.838 & 0.860 \\
Germanwings Crash & 0.867 & 0.894 \\
Ottawa Shooting & 0.871 & 0.876 \\
Sydney Siege & 0.801 & 0.829 \\
\hline
\end{tabular}%
}
\caption{Root-only classification vs.\ root--virtual link prediction (accuracy).}
\label{tab:root_vs_linkpred}
\end{table}

\vspace{-\baselineskip}
\section*{E$\quad$ Virtual-node Analysis}

We provide additional analysis of the proposed virtual-node augmentation from two perspectives. 
First, we quantify virtual-node connectivity by reporting the virtual-node degree distribution (unique neighbors) across PHEME events in Table~\ref{tab:vn_degree_stats}.
The p75 values indicate that, in typical cases, the virtual node connects to only a moderate number of nodes rather than forming a fully connected hub, while the max values highlight a small number of larger graphs where connectivity can be higher. 
Second, we examine whether the LLM-assigned subchain probability exhibits a systematic depth/length preference. 
We compute the Spearman rank correlation between node depth and the LLM-assigned probability in Table~\ref{tab:depth_bias_spearman}.
The near-zero correlations suggest that LLM scores do not consistently favor shallow or deep nodes, supporting that our virtual-edge selection is not driven by a depth bias in the LLM scoring.\\

\begin{table}[t]
\centering
\small
\setlength{\tabcolsep}{4.5pt}
\begin{tabular}{llrrr}
\hline
\textbf{Event} & \textbf{Split} & \textbf{p75} & \textbf{max} & \textbf{mean} \\
\hline
Germanwings-crash & ALL & 4 & 47 & 3.96 \\
 & Rumour & 11 & 32 & 7.81 \\
 & Non-rumour & 3 & 47 & 3.23 \\
\hline
Ferguson & ALL & 11 & 217 & 9.14 \\
 & Rumour & 18 & 94 & 13.96 \\
 & Non-rumour & 9 & 217 & 8.09 \\
\hline
Ottawa-shooting & ALL & 2 & 85 & 2.06 \\
 & Rumour & 3 & 85 & 3.71 \\
 & Non-rumour & 2 & 34 & 1.78 \\
\hline
Sydney-siege & ALL & 12 & 341 & 9.36 \\
 & Rumour & 18 & 341 & 15.32 \\
 & Non-rumour & 9 & 126 & 7.20 \\
\hline
Charliehebdo & ALL & 3 & 146 & 2.90 \\
 & Rumour & 8 & 93 & 6.40 \\
 & Non-rumour & 2 & 146 & 2.21 \\
\hline
\end{tabular}
\caption{Virtual-node degree statistics across PHEME events.}
\label{tab:vn_degree_stats}
\end{table}

\begin{table}[!ht]
\centering
\small
\setlength{\tabcolsep}{8pt}
\begin{tabular}{lc}
\hline
\textbf{Split} & \textbf{Spearman $\rho$ (depth, prob)} \\
\hline
All nodes & 0.0285 \\
Rumour (ref=0) & 0.0060 \\
Non-rumour (ref=1) & $-0.0737$ \\
\hline
\end{tabular}
\caption{Depth-bias analysis: correlation between node depth and LLM-assigned probability. Near-zero $\rho$ indicates no systematic depth/length preference in LLM scores.}
\label{tab:depth_bias_spearman}
\end{table}

\vspace{-\baselineskip}
\section*{F$\quad$Subchain Length and Token Statistics}

On large dataset Weibo, we compute the average subchain length over all graphs and all nodes. To assess worst-case input size, we additionally identify the globally longest subchain across all 4,664 Weibo event graphs and count the number of tokens after concatenating the texts of all nodes along that path. As shown in Table~\ref{tab:weibo_subchain_tokens}, subchains are typically shallow. Our input length is far below the context limits of mainstream LLM APIs, for example, DeepSeek supports 128K tokens, while GPT, Qwen, and Gemini support about 1M tokens and Grok supports 2M. Truncation is not a practical concern in our setting.\\

\begin{table}[H]
\centering
\small
\setlength{\tabcolsep}{8pt}
\begin{tabular}{lr}
\hline
\textbf{Statistic (Weibo)} & \textbf{Value} \\
\hline
Avg.\ subchain length ($\#$nodes) & 2.687 \\
Max.\ subchain length ($\#$nodes) & 73 \\
Tokens on max subchain (concat.) & 8385 \\
\hline
\end{tabular}
\caption{Weibo subchain length and token length statistics (globally longest subchain: 3907653488846171.json).}
\label{tab:weibo_subchain_tokens}
\end{table}

\FloatBarrier

\section*{G$\quad$Case Study on LLMs Performance}
\label{app:llm_case}

To illustrate the limitations of LLMs when performing rumour detection on the PHEME dataset with basic prompting, we present a representative example from the Charlie Hebdo shooting event.

\begin{itemize}
    \item \textbf{Prompt to LLM:} ``What was the Charlie Hebdo shooting?''
    
    \item \textbf{LLM Response:}
    \begin{quote}
        ``On 7 January 2015, at about 11:30 a.m. in Paris, France, the employees of the French satirical weekly magazine \textit{Charlie Hebdo} were targeted in a terrorist shooting attack by two French-born Algerian Muslim brothers, Saïd Kouachi and Chérif Kouachi. Armed with rifles and other weapons, the duo murdered 12 people and injured 11 others; they identified themselves as members of al-Qaeda in the Arabian Peninsula, which claimed responsibility for the attack\ldots''
    \end{quote}
    
    \item \textbf{Source Tweet (from PHEME dataset):}
    \begin{quote}
        ``10 people dead in Charlie Hebdo shooting in Paris.''
    \end{quote}
    
    \item \textbf{Follow-up Prompt to LLM:} ``Is this tweet spreading a rumor?''
    
    \item \textbf{LLM Response:}
    \begin{quote}
        ``No, this refers to the real Charlie Hebdo terrorist attack that occurred on 7 January 2015 in Paris.''
    \end{quote}
\end{itemize}
This case demonstrates that LLMs possess strong background knowledge of major historical events, accurately recalling key details such as the date, location, perpetrators, and official death toll (12 victims). However, when evaluating the specific tweet, the LLM overlooks a critical discrepancy: the tweet reports ``10 people dead,'' reflecting an inaccurate early report that circulated during the breaking-news phase (initial media reports often cited 10 or 11 deaths before the confirmed toll of 12).

\FloatBarrier
\section*{H$\quad$Prompt Templates for LLM-based Rumor Scoring}
\label{app:prompt_templates}

\noindent
This appendix presents the full prompt templates used in our two-stage LLM scoring procedure. We first construct a stable background knowledge base from the root post, and then use each root-to-node subchain to incrementally update the rumor probability under the same judgment basis.

\subsection*{Prompt A: Root Post Background-Knowledge Construction}
\label{app:prompt_root}

\begin{lstlisting}[basicstyle=\ttfamily\footnotesize,breaklines=true]
You are evaluating the root post of a rumor-detection thread.

Your task is to build a stable background knowledge base and a reusable judgment basis for later subchain evaluation.
Do NOT overcommit to a final rumor decision at this stage.
First identify the root post's key factual claims, then compare them against general background knowledge.
If you are uncertain, explicitly preserve uncertainty instead of inventing details.

Important principles:
(1) Separate factual assessment from stylistic suspicion.
(2) Treat direct contradiction with known facts as stronger evidence than emotional tone.
(3) Treat unsupported new key claims as moderate risk rather than conclusive proof of rumor.
(4) Emotional or forwarding-inducing language alone is only weak evidence.
(5) If the post is broadly plausible but lacks verification, do not assign an overly confident non-rumor judgment.
(6) The output must define stable and reusable criteria that can be consistently applied to later subchains.
(7) Do not assume that general familiarity with the event is sufficient evidence that the post is true; evaluate the specific claims made in the post.

Please perform the following steps:

Step 1. Extract the key claims in the root post.
For each claim, identify:
- claim_id
- claim text
- claim type: event / casualty / time / place / entity / cause / other
- verifiability: high / medium / low

Step 2. Summarize the relevant background facts that are useful for evaluating these claims.
For each fact, identify:
- fact text
- relevance: high / medium / low
- certainty: high / medium / low

Step 3. Compare each extracted claim against the background facts.
For each claim, determine:
- relation_to_background: consistent / contradicted / unsupported / uncertain
- severity: integer from 0 to 3
- note: short explanation

Severity guidance:
- 3 = direct contradiction in key factual content (e.g., entity, time, place, number, or event outcome)
- 2 = important unsupported claim or suspiciously specific new factual addition without support
- 1 = weak stylistic or rhetorical signal only
- 0 = no apparent issue

Step 4. Identify style-based risk signals without letting them dominate factual evidence.
Report:
- emotional_language: 0 or 1
- urgency_or_forwarding: 0 or 1
- missing_source_for_key_claim: 0 or 1

Step 5. Produce an initial rumor prior score for the root post.
Use a float between 0 and 1:
- 0 means strongly non-rumor
- 1 means strongly rumor
- 0.5 means uncertain / insufficient evidence

Scoring rule:
- Start from uncertainty when evidence is limited.
- Move upward mainly for factual contradiction or important unsupported claims.
- Move downward only when the post is broadly consistent with background facts and does not contain serious factual conflict.
- Avoid extreme scores unless the evidence is strong.

Return STRICT JSON ONLY. Do not use markdown fences. Do not output any extra text.

Required JSON schema:
{
  "root_claims": [
    {
      "claim_id": "c1",
      "claim": "...",
      "claim_type": "event|casualty|time|place|entity|cause|other",
      "verifiability": "high|medium|low"
    }
  ],
  "background_facts": [
    {
      "fact": "...",
      "relevance": "high|medium|low",
      "certainty": "high|medium|low"
    }
  ],
  "claim_checks": [
    {
      "claim_id": "c1",
      "relation_to_background": "consistent|contradicted|unsupported|uncertain",
      "severity": 0,
      "note": "..."
    }
  ],
  "style_risk_signals": {
    "emotional_language": 0,
    "urgency_or_forwarding": 0,
    "missing_source_for_key_claim": 0
  },
  "root_prior_rumor_score": 0.5,
  "reasoning_basis": "A concise reusable judgment basis for future subchain scoring."
}

Root post:
[ROOT_POST]
\end{lstlisting}

\subsection*{Prompt B: Subchain-based Incremental Rumor Probability Update}
\label{app:prompt_subchain}

\begin{lstlisting}[basicstyle=\ttfamily\footnotesize,breaklines=true]
You are evaluating a root-to-node subchain in a rumor-detection graph.

You must strictly follow the provided background knowledge base and reasoning basis.
Do NOT create a new evaluation standard.
Your task is to update the rumor probability of the ROOT POST using the replies in this subchain.

The subchain contains the root post followed by replies in temporal order.
Assess how the later replies affect the credibility of the root post.

Important principles:
(1) Start from the root_prior_rumor_score in the provided knowledge base.
(2) Distinguish factual correction from mere tone or agreement.
(3) Concrete evidence, contradiction, or correction matters more than emotion.
(4) Replies with no factual content should have little effect.
(5) If the evidence is mixed or weak, keep the updated score near the prior rather than becoming overconfident.
(6) Do not assume that a widely known event automatically validates the specific claims in the root post.
(7) Evaluate the replies as incremental evidence for or against the root post.

Before producing the final score, internally determine whether each non-root reply mainly functions as:
- support
- question
- deny
- correct
- add_evidence
- noise

Use these internal role judgments only to guide the probability update.
Do NOT output the role labels unless they are necessary for a brief explanation.

Update guidance:
- Increase rumor probability when a reply exposes factual inconsistency, fabricated detail, unsupported new claim, or source-related credibility problems.
- Decrease rumor probability when a reply provides concrete correction, clarification, authoritative attribution, or evidence that resolves an apparent mismatch.
- Pure agreement or pure denial without evidence should have only weak influence.
- Emotional reactions without factual content should have minimal influence.
- If the subchain adds no meaningful new evidence, the final score should remain close to the prior.

Final scoring rule:
- Begin from "root_prior_rumor_score".
- Apply only modest changes for weak or indirect evidence.
- Use larger changes only for concrete and relevant evidence.
- Keep the final score in [0, 1].
- Avoid extreme values unless the accumulated evidence is strong and consistent.

Return STRICT JSON ONLY. Do not use markdown fences. Do not output any extra text.

Required JSON schema:
{
  "probability": 0.5,
  "explanation": "A brief explanation of how the subchain updates the root rumor score."
}

Background knowledge base:
[KNOWLEDGE_BASE]

Subchain text:
[SUBCHAIN]
\end{lstlisting}

\subsection*{Design Rationale}
\label{app:prompt_rationale}

\begin{itemize}
    \item We first construct a stable prior judgment basis from the root post, then reuse it across all subchains, so that different subchains are scored under a consistent standard.
    \item We explicitly separate factual contradiction, unsupported claims, and weak stylistic signals, preventing superficial rhetoric from dominating the judgment.
    \item We model subchain scoring as an incremental update from a prior score, rather than an independent re-decision for each subchain.
    \item We require the model to internally distinguish the role of each reply (e.g., support, correction, evidence addition, or noise), but only output the final probability and a brief explanation, which reduces token cost while preserving role-aware reasoning.
    \item We preserve uncertainty when evidence is weak, improving score stability before thresholding for virtual-edge construction.
\end{itemize}

\end{document}